\documentclass[a4paper,10pt]{article}
\usepackage{amssymb}
\usepackage{amsmath}
\usepackage{graphicx}

\newcommand{\bkp}{\boldsymbol{\kappa}}

\title{The Hubbard model in strong magnetic field:\\ Low-frequency quantum oscillations due to strong electron correlations}

\author{A. Sherman\\[1ex]
{\it Institute of Physics, University of Tartu, Ravila 14c}\\
{\it 50411 Tartu, Estonia}}

\begin{document}

\maketitle

\begin{abstract}
The density of states of the two-dimensional fermionic Hubbard model in the perpendicular homogeneous magnetic field is calculated using the strong coupling diagram technique. The density of states at the Fermi level as a function of the inverse magnetic induction oscillates, and the frequency of these oscillations increases by an order of magnitude with the change of the deviation from half-filling from small to moderate values. This frequency variation is caused by the change of Landau subbands contributing to the density -- in the former case they are at the periphery of the Landau spectrum, while in the latter case the dominant contribution is provided by bands near its center. With changing induction these groups of bands behave differently. For small deviations from half-filling the calculated oscillation frequency is comparable to that observed in quantum oscillation experiments in yttrium cuprates.
\end{abstract}

\section{Introduction}
Theoretical investigations of systems of strongly correlated electrons in strong magnetic fields were started shortly after the discovery of the high-$T_c$ superconductivity. A number of works was carried out on small clusters using the exact diagonalization (see, e.g., Refs.~\cite{Castillo,Beran,Albuquerque}). It is worth noting that due to the Peierls factor \cite{Peierls}, the translation symmetry of the system is changed \cite{Brown} -- in moderate magnetic fields the size of the elementary cell increases significantly. Clusters with sizes smaller than the size of this supercell violate the symmetry of the Hamiltonian and, therefore, it is difficult to extend the obtained results to larger crystals. Another approach used for this problem is the mean-field approximation (see, e.g., Refs.~\cite{Tripathi,Yang,Schmid}). The main shortcoming of this approximation is the neglect of the dynamic character of strong correlations.

The interest in this problem was revived with the observation of low-fre\-quen\-cy quantum oscillations in the mixed state of underdoped yttrium cuprates \cite{Doiron,Bangura,Yelland,Sebastian08}. Based on the Onsager-Lifshitz-Kosevich theory for metals \cite{Shoenberg} the decreased quantum oscillation frequencies were interpreted as a manifestation of small Fermi surface pockets \cite{Sebastian12}. To explain the appearance of these small pockets proposals for various states with broken translational symmetry were suggested \cite{Millis,Chen,Galitski}. Other theories for explaining the decreased quantum oscillation frequency suppose that it is connected with superconducting fluctuations \cite{Melikyan,Pereg} or use phenomenology of the marginal Fermi liquid \cite{Varma}.

Crystals, in which the decreased quantum oscillation frequencies were observed, are underdoped cuprates, and, therefore, they are characterized by strong electron correlations. Theoretically the behavior of such crystals in strong magnetic fields is poorly known. In this work we use the strong coupling diagram technique \cite{Vladimir,Metzner,Craco,Pairault,Sherman06,Sherman15} for investigating the density of states (DOS) of the two-dimensional (2D) fermionic Hubbard model in a perpendicular homogeneous magnetic field. This approach allows us to consider large enough crystals in fields of moderate intensities. As known \cite{Brown,Langbein,Hsu}, the energy spectrum of a weakly correlated metal consists of the Landau subbands, which appear in the crystal potential instead of the Landau levels of free electrons. We found that in the approximation of a local self-energy/irreducible part each Landau subband forms strongly correlated bands independently of other subbands. Using for the irreducible part the Hubbard-I approximation \cite{Hubbard63} we revealed that the DOS at the Fermi level oscillates with changing inverse magnetic induction. The frequency $F$ of these oscillations increases by an order of magnitude when the deviation of the electron filling $\bar{n}$ from half-filling grows from a few percent to 25\%. For the Hubbard repulsion $U=8t$, $t$ being the hopping constant, and the intersite distance $a=4$\AA\ the obtained frequency is of the order of 1~kT for small values of $|1-\bar{n}|$. This frequency is close to those observed in the mentioned experiments on quantum oscillations. The origin of the strong variation of the frequency with the electron filling is in the difference in Landau subbands contributing to the DOS at the Fermi level. In the case of small deviations from half-filling these subbands are located at the periphery of the spectrum, while for larger values of $|1-\bar{n}|$ the subbands near the central part of the spectrum make the main contribution. The behavior of these two groups of subbands with the change of the magnetic induction is different. The influence of the Zeeman term on the DOS oscillations are also considered.

\section{Main formulas}
The Hamiltonian of the Hubbard model in the magnetic field reads
\begin{eqnarray}\label{Hamiltonian}
H&=&\sum_{\bf ll'\sigma}t_{\bf ll'}\exp{\left(i\frac{e}{\hbar}\int_{\bf l'}^{\bf l}{\bf A(r)}d{\bf r}\right)}a^\dagger_{\bf l'\sigma}a_{\bf l\sigma}\nonumber\\
&&+\frac{1}{2}g\mu_{\rm B}B\sum_{\bf l\sigma} \sigma a^\dagger_{\bf l\sigma}a_{\bf l\sigma}\nonumber\\
&&+\sum_{\bf l\sigma}\left(\frac{U}{2}n_{\bf l\sigma}n_{\bf l,-\sigma}-\mu n_{\bf l\sigma}\right),
\end{eqnarray}
where 2D vectors ${\bf l}$ and ${\bf l'}$ label sites of a square plane lattice, $\sigma=\pm 1$ is the projection of the hole spin, $a^\dagger_{\bf l\sigma}$ and $a_{\bf l\sigma}$ are electron creation and annihilation operators. The first, kinetic, term of the Hamiltonian contains the hopping matrix element $t_{\bf ll'}$ and the exponential factor with the Peierls phase \cite{Peierls}, in which ${\bf A}({\bf r})$ is the vector potential. The second, Zeeman, term of the Hamiltonian contains the $g$-factor $g\approx 2$, the Bohr magneton $\mu_{\rm B}$ and the magnetic induction $B$ of the external magnetic field. It is supposed that the field is directed perpendicularly to the model plane, homogeneous and is only weakly disturbed by internal currents \cite{Atkinson}. The last term of Hamiltonian (\ref{Hamiltonian}) combines the on-site Coulomb repulsion with the Hubbard constant $U$ and the electron number operator $n_{\bf l\sigma}=a^\dagger_{\bf l\sigma}a_{\bf l\sigma}$ as well as the term with the chemical potential $\mu$.

In the following consideration the Landau gauge is used, in which ${\bf A(l)}=-Bl_y{\bf x}$, where $l_y$ is the $y$ component of the site vector ${\bf l}$ and ${\bf x}$ is the unit vector along the $x$ axis. If we suppose that only the nearest neighbor hopping constant is nonzero, $t_{\bf ll'}=t\sum_{\bf a}\delta_{\bf l,l'+a}$ where ${\bf a}$ are four vectors connecting nearest neighbor sites, the Peierls exponential in the kinetic term of the Hamiltonian can be written as
\begin{equation}\label{Peierls}
{\rm e}^{i{\bf\bkp_al}},\quad \bkp_{\bf a}=\frac{e}{\hbar}Ba_x{\bf y},
\end{equation}
where $a_x$ is the $x$ component of the vector ${\bf a}$ and ${\bf y}$ is the unit vector along the $y$ axis.

Our consideration will be restricted to the fields satisfying the condition
\begin{equation}\label{condition}
\frac{e}{\hbar}Ba^2=2\pi\frac{n'}{n},
\end{equation}
where $a=|{\bf a}|$, $n$ and $n'<n$ are integers with no common factor. In this case the kinetic term of Hamiltonian (\ref{Hamiltonian}) defines its translation properties -- the Hamiltonian is invariant with respect to translations by the lattice period along the $x$ axis and by $n$ lattice periods along the $y$ axis. To retain this symmetry we apply the periodic Born-von Karman boundary conditions to the sample with $N_x$ sites along the $x$ axis and $nN_y$ sites along the $y$ axis. The boundary conditions define the set of allowed wave vectors $\left(\frac{2\pi n_x}{N_xa},\frac{2\pi n_y}{nN_ya}\right)$ with integer $n_x$ and $n_y$. As can be seen from (\ref{Peierls}) and (\ref{condition}), the momenta $\bkp_{\bf a}$ belong to this set of allowed wave vectors.

Let us consider the electron Green's function
\begin{equation}\label{Green}
G^\sigma({\bf l'\tau',l\tau})=\langle{\cal T}\bar{a}_{\bf l'\sigma}(\tau')
a_{\bf l\sigma}(\tau)\rangle,
\end{equation}
where the statistical averaging denoted by the angular brackets and time dependencies of the operators
$$a_{\bf l\sigma}(\tau)={\rm e}^{H\tau}a_{\bf l\sigma}{\rm e}^{-H\tau}\;\; {\rm and}\;\; \bar{a}_{\bf l\sigma}(\tau)={\rm e}^{H\tau}a^\dagger_{\bf l\sigma}{\rm e}^{-H\tau}$$
are determined by Hamiltonian (\ref{Hamiltonian}), ${\cal T}$ is the time-or\-de\-r\-ing operator which arranges operators from right to left in ascending order of times $\tau$. Hamiltonian (\ref{Hamiltonian}) retains the spin projection and, therefore, Green's function (\ref{Green}) is diagonal in this parameter. To calculate this function we use the strong coupling diagram technique \cite{Vladimir,Metzner,Craco,Pairault,Sherman06,Sherman15}, in which it is presented as the serial expansion with the unperturbed Hamiltonian $H_0$ given by the last term of Hamiltonian (\ref{Hamiltonian}). In the considered case the role of perturbation, over which the power expansion is carried out, is played by the first two terms of this Hamiltonian. For brevity, the sum of this two terms is denoted as
$$H_1=\sum_{\bf ll'\sigma}T^\sigma({\bf ll'})a^\dagger_{\bf l'\sigma}a_{\bf l\sigma}.$$
Terms of the serial expansion are constructed from on-site cumulants of the electron operators $a_{\bf l\sigma}(\tau)$ and $\bar{a}_{\bf l\sigma}(\tau)$ and hopping lines corresponding to the Hamiltonian $H_1$ (though the Zeeman term in $H_1$ does not lead to the transfer of an electron to another site, we retain the term ''hopping line`` used in this diagram technique). The averaging and time dependencies of operators in the cumulants are determined by the Hamiltonian $H_0$. As in the diagram technique with the expansion in powers of an interaction, in the strong coupling diagram technique the linked-cluster theorem allows one to discard disconnected diagrams and to carry out partial summations in remaining connected diagrams.

The diagram is said to be an irreducible one if it cannot be divided into two disconnected parts by cutting some hopping line. The sum of all irreducible diagrams without external ends is termed the irreducible part $K^\sigma({\bf l'\tau',l\tau})$. In terms of this quantity the equation for Green's function reads
\begin{eqnarray}\label{LarkinLT}
G^\sigma({\bf l'\tau',l\tau})&=&K^\sigma({\bf l'\tau',l\tau})+\sum_{{\bf l}_1{\bf l}'_1}\int_0^\beta K^\sigma({\bf l'\tau',l}_1\tau_1)\nonumber\\
&&\times T^\sigma({\bf l}_1{\bf l}'_1)G^\sigma({\bf l}'_1\tau_1,{\bf l\tau})d\tau_1,
\end{eqnarray}
or after the Fourier transformations over space and time variables
\begin{eqnarray}\label{LarkinQM}
G^\sigma({\bf q'q}m)&=&K^\sigma({\bf q'q}m)+\sum_{{\bf q}_1{\bf q}'_1} K^\sigma({\bf q'q}_1m)\nonumber\\
&&\times T^\sigma({\bf q}_1{\bf q}'_1)G^\sigma({\bf q}'_1{\bf q}m),
\end{eqnarray}
where $\beta=1/T$ is the inverse temperature, $m$ is an integer determining the Matsubara frequency $\omega_m=(2m+1)\pi T$ and
\begin{eqnarray}\label{hopping}
T^\sigma({\bf qq}')&=&t\Big[{\rm e}^{iq'_xa}\delta_{{\bf q,q'+\bkp}n'}+{\rm e}^{-iq'_xa}\delta_{{\bf q,q'-\bkp}n'}\nonumber\\
&&+2\cos(q'_ya)\delta_{\bf qq'}\Big]
+\frac{1}{2}g\mu_{\rm B}B\sigma\delta_{\bf qq'},
\end{eqnarray}
$q_x$ and $q_y$ are components of the wave vector ${\bf q}$, which belong to the first Brillouin zone, and $\bkp=\frac{2\pi}{na}{\bf y}$. In the derivation of (\ref{hopping}) we took into account that $\bkp_{\bf a}$ in (\ref{Peierls}) belongs to the set of momenta determined by the chosen periodic boundary conditions. Equations (\ref{LarkinLT}) and (\ref{LarkinQM}) are forms of the Larkin equation \cite{Larkin}. Analogous equations were used for calculating the electron Green's function of the Hubbard model in the absence of the magnetic field \cite{Vladimir,Metzner,Craco,Pairault,Sherman06,Sherman15}.

In (\ref{LarkinQM}), $G^\sigma({\bf q'q}m)$ and $K^\sigma({\bf q'q}m)$ are not diagonal with respect to momenta due to the reduced translation symmetry of the problem. However, from symmetry arguments one can see that these quantities are nonzero only for ${\bf q'=q+\nu\bkp}$ with an integer $\nu$ [for a momentum independent irreducible part or, equivalently, self-energy this statement follows directly from (\ref{LarkinQM}) and (\ref{hopping})]. In view of this result it is convenient to split the first Brillouin zone into $n$ streaks of the width $\frac{2\pi}{na}$, which are oriented parallel to the $x$ axis. Let us denote wave vectors in the streak with $-\frac{\pi}{a}<q_y\leq -\frac{\pi}{a}+\frac{2\pi}{na}$ and $-\frac{\pi}{a}<q_x\leq\frac{\pi}{a}$ as ${\bf k}$. Then any wave vector in the first Brillouin zone can be represented as ${\bf k}+j\bkp$ with $0\leq j\leq n-1$. In these notations
$$G^\sigma({\bf q'q}m)=G^\sigma({\bf k}+j'\bkp,{\bf k}+j\bkp,m)\equiv G^\sigma_{j'j}({\bf k}m).$$
Analogous notations can be used for $K^\sigma({\bf q'q}m)$ and $T^\sigma({\bf q'q})$. Considering indices $j$ and $j'$ as matrix indices equation (\ref{LarkinQM}) can be rewritten as
\begin{equation}\label{LarkinKM}
{\bf G}^\sigma({\bf k}m)=\left[{\bf 1}-{\bf K}^\sigma({\bf k}m){\bf T}^\sigma({\bf k})\right]^{-1}{\bf K}^\sigma({\bf k}m),
\end{equation}
where matrices are denoted by the boldface font and ${\bf 1}$ is a $n\times n$ identity matrix.

If the approximation of the dynamic mean field the\-ory \cite{Georges} -- a momentum independent self-energy -- is accepted, Eq. (\ref{LarkinKM}) acquires the form
\begin{equation}\label{LarkinKW}
{\bf G}^\sigma({\bf k}\omega)=\left\{\left[K^\sigma(\omega)\right]^{-1}{\bf 1}-{\bf T}^\sigma({\bf k})\right\}^{-1},
\end{equation}
where the analytic continuation to the real frequency axis was performed. In this approximation each Landau subband forms strongly correlated bands independently of other subbands. Indeed, let us denote the part of ${\bf T}^\sigma({\bf k})$, Eq. (\ref{hopping}), which is proportional to $t$, as ${\bf T'(k)}$. It is a Hermitian matrix with the eigenvectors ${\bf V_\lambda(k)}$ and eigenvalues $E_\lambda({\bf k})$,
\begin{equation}\label{Harper}
{\bf T'(k)V_\lambda(k)}=E_\lambda({\bf k}){\bf V_\lambda(k)},\quad 0\leq\lambda\leq n-1.
\end{equation}
This equation is the Harper equation \cite{Harper} for calculating dispersions $E_\lambda({\bf k})$ of the Landau subbands. The vectors ${\bf V_\lambda(k)}$ are also the eigenvectors of the matrix ${\bf G}^\sigma({\bf k}\omega)$,
\begin{eqnarray}\label{poles}
{\bf G}^\sigma({\bf k}\omega){\bf V_\lambda(k)}&=&\left\{\left[K^\sigma(\omega)\right]^{-1}- E_\lambda({\bf k})-\frac{1}{2}g\mu_{\rm B}B\sigma\right\}^{-1}\nonumber\\
&&\times{\bf V_\lambda(k)}
\end{eqnarray}
Except the Zeeman contribution the braces in the above equation look like the poles of the electron Green's function of the Hubbard model with the initial dispersion $E_\lambda({\bf k})$ in zero field \cite{Sherman06,Sherman15}. Hence each Landau subband $\lambda$ forms strongly correlated bands independently of other subbands. In other words, with respect to strong correlations the Landau subbands behave as independent bands in this approximation. Notice also that for vanishing $U$ only the first-order cumulant remains nonzero, $\left[K^\sigma(\omega)\right]^{-1} \rightarrow \omega-\mu$ and the above equations reduce to Green's function of uncorrelated electrons in the magnetic field.

From Eq. (\ref{poles}) we find for the DOS
\begin{eqnarray}\label{DOS}
\rho^\sigma(\omega)&=&-\frac{1}{\pi N}{\rm Im}\sum_{\bf k}{\rm Tr}{\bf G}^\sigma({\bf k\omega})\nonumber\\
&=&-\frac{1}{\pi N}{\rm Im}\sum_{\bf k\lambda}\bigg\{\left[K^\sigma(\omega)\right]^{-1}- E_\lambda({\bf k})\nonumber\\
&&-\frac{1}{2}g\mu_{\rm B}B\sigma\bigg\}^{-1},
\end{eqnarray}
where $N=nN_xN_y$ is the number of sites.

In the below calculations the irreducible part $K^\sigma(\omega)$ was approximated by the first-order cumulant $C_1(\tau'\tau)=\langle{\cal T}\bar{a}_{\bf l\sigma}(\tau')a_{\bf l\sigma}(\tau)\rangle_0$, where the subscript 0 indicates that the averaging and time dependencies are determined by $H_0$. In the absence of the magnetic field this approximation leads to the Hubbard-I approximation \cite{Vladimir}. For the chemical potential satisfying the conditions
\begin{equation}\label{conditions}
T\ll\mu,\quad T\ll U-\mu
\end{equation}
the cumulant reads
\begin{equation}\label{C1}
C_1(\omega)=\frac{\omega+\mu+\frac{U}{2}}{(\omega+\mu)(\omega+\mu-U)}.
\end{equation}
Substituting this approximation for the irreducible part into (\ref{DOS}) we find
\begin{eqnarray}\label{DOSf}
\rho^\sigma(\omega)&=&\frac{1}{2N}\sum_{\bf k\lambda}\frac{1}{\sqrt{U^2+{\cal E}^2({\bf k\lambda\sigma})}}\nonumber\\
&&\times\Big\{\left[\sqrt{U^2+{\cal E}^2({\bf k\lambda\sigma})}+{\cal E}({\bf k\lambda\sigma})\right]\delta\left(\omega-\varepsilon^\sigma_{\bf k\lambda+}\right)\nonumber\\
&&\;\;+\left[\sqrt{U^2+{\cal E}^2({\bf k\lambda\sigma})}-{\cal E}({\bf k\lambda\sigma})\right]\nonumber\\
&&\;\;\times\delta\left(\omega-\varepsilon^\sigma_{\bf k\lambda-}\right)\Big\},
\end{eqnarray}
where
\begin{eqnarray}\label{energies}
&&{\cal E}({\bf k\lambda\sigma})=E_\lambda({\bf k})+\frac{1}{2}g\mu_{\rm B}B\sigma,
\nonumber\\[-1ex]
&&\\[-1ex]
&&\varepsilon^\sigma_{\bf k\lambda\pm}=\frac{1}{2}\left[U+{\cal E}({\bf k\lambda\sigma})\right]\pm \frac{1}{2}\sqrt{U^2+{\cal E}^2({\bf k\lambda\sigma})}-\mu\nonumber
\end{eqnarray}
and $E_\lambda({\bf k})$ is a solution of the Harper equation (\ref{Harper}) with the tridiagonal cyclic matrix ${\bf T'(k)}$ from (\ref{hopping}).

\section{Results and discussion}
Below we consider the density of electron states (\ref{DOSf}). For $T=0$ the integral of the product of this quantity and the frequency over the occupied states gives the electron contribution to the thermodynamic potential $\Omega$. As in the case of weak electron correlations, oscillations of $\rho(\omega=0)$, the DOS at the Fermi level, in varying magnetic field lead to oscillations in $\Omega$ and its derivatives, which are observed in quantum oscillation measurements \cite{Shoenberg}.

\begin{figure}[t]
\centerline{\resizebox{0.9\columnwidth}{!}{\includegraphics{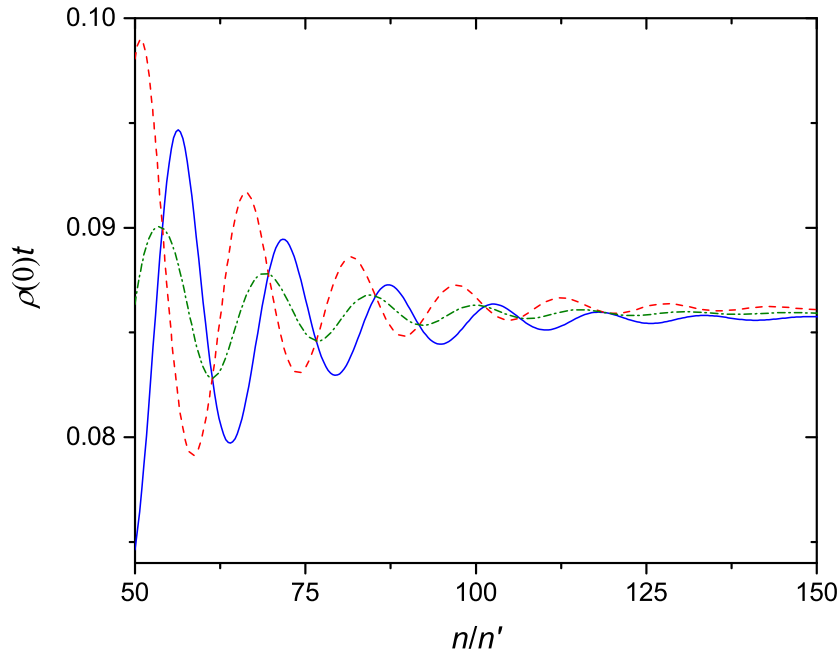}}}
\caption{(Color online) The densities of states at the Fermi level $\rho^{(-1)}$ (the blue solid line), $\rho^{(+1)}$ (the red dashed line) and their mean value $\frac{1}{2}\left(\rho^{(-1)}+\rho^{(+1)}\right)$ (the olive dash-dotted line) as functions of the inverse magnetic induction expressed in terms of $\frac{n}{n'}=\frac{h}{ea^2}\frac{1}{B}$. $T=0$, $U=8t$, $n'=3$, $\bar{n}=0.953$ and $\frac{g\mu_{\rm B}h}{2ea^2}=2t$.}\label{Fig1}
\end{figure}
Let us first consider the difference in the oscillations of $\rho^\sigma(0)$ for electrons, which spins are parallel and antiparallel to the applied magnetic field. The difference is caused by the Zeeman term of Hamiltonian (\ref{Hamiltonian}). An example is given in Fig.~\ref{Fig1}. In these calculations we have substituted $\delta$-functions in (\ref{DOSf}) with Lorentzians with the artificial broadening $\eta=0.03t$, which imitates finite lifetimes of states. The oscillations are very sensitive to this broadening: with increasing $\eta$ the oscillation amplitude decreases rapidly. The parameter $z=\frac{g\mu_{\rm B}h}{2ea^2}$, indicated in the figure caption, characterizes the value of the Zeeman term. For $a=4$\AA, the intersite distance approximately corresponding to the distance between Cu ions in Cu-O planes of YBa$_2$Cu$_3$O$_{7-y}$, this parameter is equal to 1.5 eV. For zero temperature the electron concentration is calculated using the formula
\begin{equation}\label{concentration}
\bar{n}=\sum_\sigma\int_{-\infty}^0\rho^\sigma(\omega)d\omega.
\end{equation}
As seen from the figure, the oscillations in $\rho^{(-1)}(0)$ and $\rho^{(+1)}(0)$ are shifted in phase, and this shift grows with growing $z/t$. However, the frequencies of the two oscillations are the same and equal to $F\approx 1.7$~kT. Notice that this frequency by an order of magnitude smaller than oscillation frequencies in metals with large Fermi surfaces \cite{Shoenberg}. The frequency does not change with the variation of $B$ or $z$. Due to the phase shift the amplitude of oscillations in $\bar{\rho}=\frac{1}{2}\left(\rho^{(-1)}+\rho^{(+1)}\right)$, which enters into $\Omega$, is smaller than in any of the two summands. However, even in the worst case when oscillations in them are in antiphase, as in Fig.~\ref{Fig1}, the complete compensation of oscillations does not occur in $\bar{\rho}$.

\begin{figure}[t]
\centerline{\resizebox{0.75\columnwidth}{!}{\includegraphics{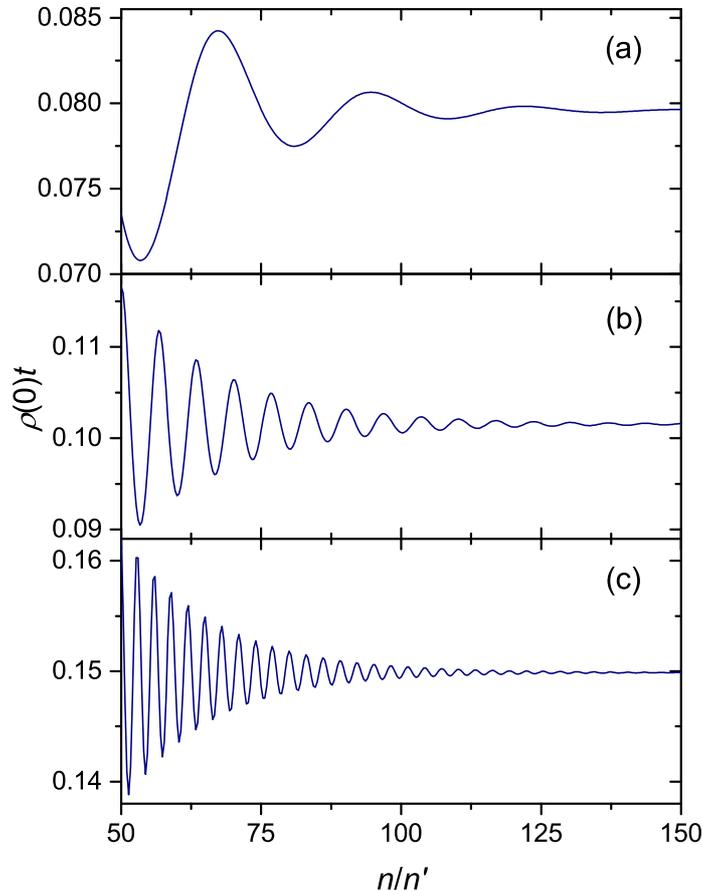}}}
\caption{(Color online) The density of states at the Fermi level as a function of the inverse magnetic induction expressed in terms of $\frac{n}{n'}=\frac{h}{ea^2}\frac{1}{B}$ for $\bar{n}=0.969$ (a, $F=0.95$~kT), $\bar{n}=0.897$ (b, $F=3.9$~kT) and $\bar{n}=0.751$ (c, $F=8.6$~kT). $T=0$, $U=8t$ and $n'=3$.}\label{Fig2}
\end{figure}
As follows from the symmetry of the Hamiltonian, for $\omega=0$
$$\rho^\sigma(t,\mu)=\rho^{-\sigma}(t,U-\mu)=\rho^\sigma(-t,\mu).$$
Therefore, below we set $t>0$ and $\mu<\frac{U}{2}$, which corresponds to $\bar{n}<1$. Besides, we neglect the Zeeman term, which does not qualitatively change results discussed below. The variation of the DOS oscillation with the electron filling is shown in Fig.~\ref{Fig2}. In the figure caption, the oscillation frequencies $F$ are estimated by setting $a=4$\AA, as above. As seen from the figure, the frequency $F$ is changed by an order of magnitude when the electron filling is decreased from $\bar{n}=0.969$ to $0.751$. This corresponds to the shift of the chemical potential from the value $\mu=1.4t$ near the top of the lower Hubbard subband to $\mu=0.4t$.

What is the reason for small values of $F$ at small deviations from half-filling? As was indicated in the previous section, a Landau subband forms strongly correlated bands independently of other subbands. From (\ref{energies}) we find for the energies of Landau subbands contributing to the DOS at the Fermi level
\begin{equation}\label{torho0}
E_\lambda({\bf k})\approx\frac{\mu(U-\mu)}{\frac{U}{2}-\mu}.
\end{equation}
This equation is obtained for the case $\bar{n}<1$ ignoring the Zeeman contribution. Notice that for $\mu\rightarrow 0$, when the chemical potential goes deep into the lower Hubbard subband and the value $1-\bar{n}$ is large, $E_\lambda({\bf k})\rightarrow 0$. Landau subbands $E_\lambda({\bf k})$ obtained from the Harper equation (\ref{Harper}) are located in the energy range $\left(-\frac{\Delta}{2},\frac{\Delta}{2}\right)$, where $\Delta=8t$ is the width of the initial band. The above result means that for large deviations from half-filling Landau subbands near the center of this range contribute to $\rho(0)$. In this spectral region the number of Landau subbands is large and their positions change rapidly with the variation of $B$ \cite{Langbein,Hsu}. This leads to high-frequency oscillations of $\rho(0)$ with changing induction for such electron concentrations. On the other hand, the value of the chemical potential near the top of the lower Hubbard subband, when $1-\bar{n}$ is small, can be also estimated from (\ref{energies}),
$$\mu_{\rm top}\approx\frac{1}{2}\left(U+\frac{\Delta}{2}\right)-\frac{1}{2}\sqrt{U^2+ \frac{\Delta^2}{4}}.$$
Substituting this result into (\ref{torho0}) we get $E_\lambda({\bf k})\approx\frac{\Delta}{2}$. In other words, in this case Landau subbands at the periphery of their energy range contribute to $\rho(0)$. The number of these subbands is smaller and the distance between them is larger than in the central part of the Landau spectrum \cite{Langbein,Hsu}. This results in the slower frequency of oscillations observed in $\rho(0)$ for these $\mu$.

The above discussion is valid for the case $U\gtrsim\Delta$. Though the Hubbard-I approximation used above is rather rough for smaller $U$, we hope it gives at least qualitatively correct results. For $U\ll\Delta$ widths of the bands $\varepsilon_{\bf k\lambda\pm}$, Eq.~(\ref{energies}),  become larger, and a larger number of the Landau subbands contributes to $\rho(0)$. As a result, in contrast to the case of strong correlations, for $\mu\approx\mu_{\rm top}$ the main contribution to $\rho(0)$ is made by subbands, which are well off the spectrum edges $\pm\frac{\Delta}{2}$. Therefore, for $U\ll\Delta$ the variation of the oscillation frequency with the change of the electron filling is less pronounced in comparison with the case $U\gtrsim\Delta$, and the oscillations are observable only for large field intensities.

For $U\gtrsim\Delta$ the frequencies $F$ obtained above for small deviations from half-filling are of the same order of magnitude as those found in lightly doped cuprates \cite{Doiron,Bangura,Yelland,Sebastian08}. The mechanism, which leads to these decreased frequencies, is based on the important feature of these crystals -- strong electron correlations. This gives promise that the mechanism can be used for the interpretation of the mentioned experiments on quantum oscillations.

\section{Concluding remarks}
In this work the two-dimensional fermionic Hubbard model in a perpendicular homogeneous magnetic field was considered. Using the strong coupling diagram technique expressions for the electron Green's function and the density of states were derived. Calculations were performed for the nearest neighbor form of the kinetic energy and for the approximation of the irreducible part, which corresponds to the Hubbard-I approximation. For a momentum independent irreducible part/self-energy each Landau subband was found to form strongly correlated bands independently of other subbands. The density of states at the Fermi level as a function of the inverse magnetic induction demonstrates oscillations, which lead to oscillations in the thermodynamic potential and its derivatives, observed in quantum oscillation measurements. Due to the Zeeman term the densities of states of the electrons with spins parallel and antiparallel to the magnetic field are shifted in phase. However, the full compensation of these oscillations in the combined density of states does not occur. For Hubbard repulsions comparable to the width of the initial electron band frequencies of oscillations for small deviations from half-filling are much smaller than those for larger deviations. The reason of this is the fact that for different electron fillings different parts of the spectrum of Landau subbands contribute to the density of states at the Fermi level. For small deviations from half-filling it is the periphery of this spectrum, while for larger deviations it is a part near the center of the spectrum, which varies faster in changing field than the periphery. Oscillation frequencies for small deviations from half-filling are comparable with those observed in quantum oscillation experiments on lightly doped yttrium cuprates.

This work was supported by the research project IUT2-27, the European Regional Development Fund TK114 and the Estonian Scientific Foundation (grant ETF9371).


\begin{thebibliography}{99}
\bibitem{Castillo}H.E. Castillo and C.A. Balseiro, Phys.\ Rev.\ Lett.\ {\bf 68}, 121 (1992).
\bibitem{Beran}P. B\'eran, Phys.\ Rev.\ B {\bf 54}, 1391 (1996).
\bibitem{Albuquerque}A.F. Albuquerque and G.B. Martins, J.\ Phys.: Condens.\ Matter {\bf 17}, 2419  (2005).
\bibitem{Peierls}R. Peierls, Z.\ Phys.\ {\bf 80}, 763 (1933).
\bibitem{Brown}E. Brown, Phys.\ Rev.\ {\bf 133}, A1038 (1964).
\bibitem{Tripathi}G.S. Tripathi, Phys.\ Rev.\ B {\bf 52}, 6522 (1995).
\bibitem{Yang}Yong Wang and A.H. MacDonald, Phys.\ Rev.\ B {\bf 52}, R3876 (1995).
\bibitem{Schmid}M. Schmid, B.M. Andersen, A.P. Kampf, and P.J. Hirsch\-feld, New J.\ Phys.\ {\bf 12}, 053043 (2010).
\bibitem{Doiron}N. Doiron-Leyraud, C. Proust, D. LeBoeuf, J. Levallois, J.-B. Bonnemaison, Ruixing Liang, D.A. Bonn, W.N. Hardy, and L. Taillefer, Nature {\bf 447}, 565 (2007).
\bibitem{Bangura}A.F. Bangura, P.A. Goddard, J. Singleton, S.W. Tozer, A.I. Coldea, A. Ardavan, R.D. McDonald, S.J. Blundell, and J.A. Schlueter, Phys.\ Rev.\ B {\bf 76}, 052510 (2007).
\bibitem{Yelland}E.A. Yelland, J. Singleton, C.H. Mielke, N. Harrison, F.F. Balakirev, B. Dabrowski, and J.R. Cooper, Phys.\ Rev.\ Lett.\ {\bf 100}, 047003 (2008).
\bibitem{Sebastian08}S.E. Sebastian, N. Harrison, E. Palm, T.P. Murphy, C.H. Mielke, Ruixing Liang, D.A. Bonn, W.N. Hardy, and G.G. Lonzarich, Nature {\bf 454}, 200 (2008).
\bibitem{Shoenberg}D. Shoenberg, {\it Magnetic oscillations in metals} (Cambridge University Press, Cambridge, 1984).
\bibitem{Sebastian12}S.E. Sebastian, N. Harrison, and G.G. Lonzarich, Rep.\ Prog.\ Phys.\ {\bf 75}, 102501 (2012).
\bibitem{Millis}A.J. Millis and M.R. Norman, Phys.\ Rev.\ B {\bf 76}, 220503(R) (2007).
\bibitem{Chen}W.-Q. Chen, K.-Y. Yang, T.M. Rice, and F.C. Zhang, Europhys.\ Lett.\ {\bf 82}, 17004 (2008).
\bibitem{Galitski}V. Galitski and S. Sachdev, Phys.\ Rev.\ B {\bf 79}, 134512 (2009).
\bibitem{Melikyan}A. Melikyan and O. Vafek, Phys.\ Rev.\ B {\bf 78}, 020502(R) (2008).
\bibitem{Pereg}T. Pereg-Barnea, H. Weber, G. Rafael, and M. Franz, Nature Phys.\ {\bf 6}, 44 (2010).
\bibitem{Varma}C.M. Varma, Phys.\ Rev.\ B {\bf 79}, 085110 (2009).
\bibitem{Vladimir}M.I.~Vladimir and V.A.~Moskalenko, Teor.\ Mat.\ Fiz.\ {\bf 82}, 428 (1990) [Theor.\ Math.\ Phys.\ {\bf 82}, 301 (1990)]; S.I.~Vakaru, M.I.~Vladimir, and V.A.~Moskalenko, Teor.\ Mat.\ Fiz.\ {\bf 85}, 248 (1990) [Theor.\ Math.\ Phys.\ {\bf 85}, 1185 (1990)]; V.A.~Moskalenko, P.~Entel, and D.F.~Digor, Phys.\ Rev.\ B {\bf 59}, 619 (1999).
\bibitem{Metzner}W.~Metzner, Phys.\ Rev.\ B {\bf 43}, 8549 (1991).
\bibitem{Craco}L.~Craco and M.A.~Gusm\~{a}o, Phys.\ Rev.\ B {bf 52}, 17135 (1995); {\bf 54}, 1629 (1996); L.~Craco, J.\ Phys.: Condens. Matter {\bf 13}, 263 (2001).
\bibitem{Pairault}S.~Pairault, D.~S\'en\'echal, and A.-M.S.~Tremblay, Eur.\ Phys.\ J.\ B {\bf 16}, 85 (2000).
\bibitem{Sherman06}A.~Sherman, Phys.\ Rev.\ B {\bf 73}, 155105 (2006); {\bf 74}, 035104 (2006).
\bibitem{Sherman15}A.~Sherman, Physica B {\bf 456}, 35 (2015); Int.\ J.\ Mod. Phys. B {\bf 29}, 1550088 (2015); arXiv:1501.03587.
\bibitem{Langbein}D. Langbein, Phys.\ Rev.\ {\bf 180}, 633 (1969).
\bibitem{Hsu}W.Y. Hsu and L.M. Falicov, Phys.\ Rev.\ B {\bf 13}, 1595 (1976).
\bibitem{Hubbard63}J.~Hubbard, Proc.\ R.\ Soc.\ Lond.\ A {\bf 276}, 238 (1963); {\bf 277}, 237 (1964).
\bibitem{Atkinson}W.A. Atkinson and J.E. Sonier, Phys.\ Rev.\ B {\bf 77}, 024514 (2008).
\bibitem{Larkin}A.I. Larkin, Zh. Eksp. Teor. Fiz. {\bf 37}, 264 (1959) [Sov. Phys. JETP {\bf 37}, 186 (1960)].
\bibitem{Georges}A.~Georges, G.~Kotliar, W.~Krauth, and M.~Rozenberg, Revs.\ Mod.\ Phys.\ {\bf 68}, 13 (1996).
\bibitem{Harper}P.G.~Harper, Proc.\ Phys.\ Soc., Lond.\ A {\bf 68}, 874 (1955).
\end{thebibliography}
\end{document}